\documentclass{JHEP}
\usepackage{latexsym}

\makeatletter

\newcommand{\nn}{\nonumber\\}\newcommand{\p}[1]{(\ref{#1})}

\newcommand{\AmS}{{\protect\the\textfont2
  A\kern-.1667em\lower.5ex\hbox{M}\kern-.125emS}}
\makeatletter
 \@addtoreset{equation}{section}

 \makeatother

\preprint{Preprint MRI-P-030101, DFPD 03/TH/01}

\title{Supersymmetric and $\kappa$--invariant Coincident D0--Branes}

\author{Sudhakar Panda $^a$\thanks{panda@mri.ernet.in} ~and Dmitri Sorokin
$^{b,\,c}$ \thanks{sorokin@pd.infn.it}
\\
~\\
 ${}^a$ {\it\small Harish-Chandra Research Institute}\\
{\it\small Chhatnag Road, Jhushi, Allahabad, 211019, India}\\
~\\
 ${}^b$ {\it\small Institute for Theoretical Physics, NSC KIPT}\\
{\it\small Akademicheskaya Street 1, 61108 Kharkov, Ukraine}\\
~\\
 ${}^c$
{\it\small Universit\`a Degli Studi di Padova,
Dipartimento di Fisica ``Galileo Galilei''}\\
{\it\small ed INFN, Sezione Di Padova Via F. Marzolo, 8, 35131
Padova, Italia}
 }
\date{}

\abstract{We propose a generic supersymmetric and
$\kappa$--invariant action for describing coincident D0--branes
with non--abelian matter fields on their worldline. The action is
shown to be in agreement with the Matrix Theory limit of the
ND0--brane effective action.}

\begin{document}

\section{Introduction}
The construction of a complete action for the supersymmetric
non--abelian Dirac--Born--Infeld theory which could be associated
with the effective action of the open superstring theory, and the
construction of corresponding supersymmetric actions for
coincident Dirichlet branes is still lacking. At the bosonic level
the problem is that though one knows the full structure of all the
non--abelian terms which enter the effective action under the
symmetrized trace in a DBI manner
\cite{Tseytlin:1999dj,Myers:1999ps,Taylor:2000pr}, a generic
recipe of constructing the commutator and higher derivative terms
\cite{roo} has not been found yet. Also adding fermionic terms to
the symmetrized trace part of the non-abelian DBI action for
coincident branes via supersymmetrization encounters difficulties.
By now the supersymmetric and $\kappa$--invariant actions with the
symmetrized trace have been constructed only for N coincident
D0--branes (ND0--branes) in $N=1$, $D=2$ \cite{Sorokin:2001av},
with partial results obtained also for super ND0--branes in a
space--time of arbitrary dimension \cite{partial}, and for space
filling ND2--branes in $N=2$, $D=3$ superspace \cite{dhl}. The
space filling supersymmetric ND2--brane model of \cite{dhl} has
been constructed with the use of superembedding methods (see
\cite{super} for a review), while the super ND0--brane models have
been obtained by more traditional methods starting from a
reparametrization invariant form of the bosonic action for N
coincident D0--branes and making it supersymmetric and
$\kappa$--invariant. We shall follow the point of view of above
mentioned papers that N coincident D--branes should, actually, be
regarded as a single brane with non--abelian fields on its
worldvolume, or a N(onabelian)D--brane. The worldvolume
diffeomorphisms and $\kappa$--symmetry of the ND--brane to be
conventional ``abelian'' symmetries in contrast to the
``non--abelian" proposal of \cite{b}. A main problem of
generalizing these results to higher dimensions is to find an
explicit form of the first--class constraint which is responsible
for the worldline reparametrization invariance of the non--abelian
ND0--brane actions. In a recent paper \cite{pp} this problem has
been solved for the case of ND0--branes in $D=3$ and a worldline
reparametrization invariant bosonic ND0--brane action of
codimension two has been constructed.

The purpose of this paper is to generalize the results of
\cite{Sorokin:2001av,partial} and \cite{pp} to supersymmetric N
coincident D0--branes in higher dimensional superspaces. The main
result we get is a generic  supersymmetric and $\kappa$--invariant
action for describing coincident D0--branes with non--abelian
matter fields on their worldline. Particular examples to be
considered are the non-abelian ND0--branes in $N=2$, $D=3$
superspace and in type IIA $D=10$ superspace. The action is
demonstrated to be in agreement with the Matrix Theory limit of
the ND0--brane effective action \cite{bfss}.

\section{Reparametrization invariant bosonic ND0--brane action}
For the purposes of supersymmetrization it turns out to be
convenient to use the first order formulation of the ND0--brane
action in a D--dimensional Minkowski space parametrized by {\sl
commuting} coordinates $x^M=(x^0,x^i)$ $(i=1,\cdots,D-1)$
\cite{Sorokin:2001av}
\begin{equation}\label{1}
S=\int d\tau\, {\rm Tr}\left\{{1\over N}p_M\dot
x^M+p_{\varphi^i}\dot\varphi^i\right. \left. -{{e(\tau)}\over
2N}[p_Mp^M+M^2(p_i,p_\varphi,\varphi^j)]\right\}\,,
\end{equation}
where $e(\tau)$ is the Lagrange multiplier for the mass shell
condition and $M(p_{\varphi^i},p_i,\varphi^j)$ is an `effective
mass' which is an ordinary constant mass in the case of a single
D0--brane, and which in the case of N coincident D0-branes depends
on the $U(N)$--valued canonical conjugate momenta
$P_{\Phi^i}={1\over N}\,p_i\,I+p_{\varphi^i}$ of the non--abelian
$U(N)$ scalar fields $\Phi^i(\tau)=x^i(\tau)\,I+\varphi^i(\tau)$.
$x^M=(x^0,x^i)$ are the center of mass space--time coordinates of
the ND0--brane and $\varphi^i(\tau)$ are $SU(N)$ valued worldline
fields, traceless hermitian $N\times N$ matrices.

In the case of the ND0--brane in $D=2$ the effective mass does not
depend on $\varphi^i(\tau)$ and has the form \cite{Sorokin:2001av}
\begin{equation}\label{2}
M^2=\left[\left({\rm Tr} \sqrt{({1\over
N}p_1+p_\varphi)^2+m^2}\right)^2-p_1^2\right]\,.
\end{equation}
In the above, $m$ is the mass (or `tension') of a single D0 brane.

In the $D=3$ case the effective mass acquires a dependence on
$\varphi^i(\tau)$ and becomes \cite{pp}
\begin{eqnarray}\label{3}
M^2(p_i,p_{\varphi^i},\varphi^j) &=& \left[\left({\rm STr}
\sqrt{(p^2_{\Phi^1} + p^2_{\Phi^2} + m^2) (1 - [\Phi^1,
\Phi^2]^2)}\right)^2 -p^2_1 -p^2_2\right]\, \nonumber\\
&=&\left[\left({\rm STr} \sqrt{[(p_{\Phi^i})^2  + m^2 ]\, det\,
Q^{ij}}\right)^2 -p^2_i \right]\,,
 \end{eqnarray}
where the trace is symmetrized, $Q=1$ for $D=2$ and
$Q^{ij}=\delta^{ij}+i[\Phi^i,\Phi^j]=\delta^{ij}+i[\phi^i,\phi^j]$
$(i,j=1,2,\cdots, D-1)$ for $D=3$.

Using the explicit form of $M$ in $D=2,3$, integrating \p{1} over
the canonical momenta and imposing the static gauge $x^0=\tau$ one
gets the non--abelian DBI--like ND0--brane action of
\cite{Tseytlin:1999dj,Myers:1999ps,Taylor:2000pr} in a flat
background
\begin{equation}\label{3.1}
{S}=-m\, \int d\tau\,{\rm STr}\left\{ \sqrt{\left(
1-\dot\Phi^{i}\,Q^{-1}_{ij}\,\dot\Phi^{j}\right)\,det\,Q^{ij}}\right\}\,.
\end{equation}

In higher dimensions the expression for the effective mass $M$
should be a generalization of \p{3}, but its explicit complete
form  has not been found yet because of technical problems.

 However, as we will see in the next section, it
turns out that the explicit form of $M$ does not play any
important role for the construction of a supersymmetric action for
ND0--branes which happens to be consistent for a generic form of
the effective mass.

\section{Supersymmetrization} We shall now generalize the action
(\ref{1}) to be supersymmetric and $\kappa$--invariant in a
D--dimensional $N=2$ superspace whose Grassmann directions are
parametrized by {\sl anticommuting} spinor coordinates
$\theta^\alpha$, $\alpha$ being a commulative index which
 stands for a spinorial index of a corresponding $Spin(1,D-1)$
 group and can also include an $SO(2)$ R--symmetry index $I=1,2$ of $N=2$
 supersymmetry, when present, as in the case $N=2$, $D=3$. $\theta^\alpha(\tau)$
 are Grassmann--odd coordinates in the superspace
 of the center of mass of the N coincident D0--branes.

A generic form of the supersymmetric ND0--brane action is
\cite{Sorokin:2001av,partial}
%\begin{equation}
\begin{eqnarray}\label{4}
 S=\int d\tau {\rm Tr}\left\{{1\over N}p_M(\dot
x^M+i\bar\theta\gamma^M\dot\theta)\right. -{{e(\tau)}\over
2N}\left.\left[p_Mp^M+M^2(p_i,p_{\varphi},\varphi,\psi)\right]\right.
&&\nn
 \left.+{i\over N}\,M(p_i,p_{\varphi},\varphi,\psi)
\,\bar\theta\Gamma\dot\theta \right. \left.
+p_{\varphi^i}\dot\varphi^i+i\bar\psi^A\dot\psi^A\right\}\,,
\qquad\,\qquad\,\qquad\,\qquad\,
%\end{equation}
\end{eqnarray}
where in the `Chern--Simons' part of the ND0--brane action
$\Gamma$ is a spinor matrix with the properties that
$\Gamma\Gamma=I, \{\Gamma,\,\gamma^m\}=0$ and whose explicit form
depends on the dimension of the target superspace. For instance,
in $D=2$ $\theta^\alpha$ ($\alpha$=1,2) is a two component
Majorana spinor and $\Gamma=\gamma^2$ where,
$\gamma^2=\gamma^0\,\gamma^1$ are $D=2$ Dirac matrices. In $N=2$,
$D=3$ we have two two--component Majorana spinors
$\theta^{I\alpha}$ $(I=1,2)$,
$\gamma^M_{I\alpha,J\beta}=\delta_{IJ}\,\gamma^M_{\alpha\beta}$
and $\Gamma_{I\alpha,J\beta}=\epsilon_{IJ}\epsilon_{\alpha\beta}$
where $\epsilon$ are $2\times 2$ antisymmetric unit matrices. In
type IIA $D=10$ superspace $\theta^\alpha$ is a 32--component
Majorana spinor and  $\Gamma=\gamma^{11}$.

$\psi^A(\tau)$ appearing in the action \p{4} are Grassmann--odd
$SU(N)$--valued worldline fields.  Note that we assume that the
effective mass of the supersymmetric ND0--brane \p{4} can also
depend on the fields $\psi^A(\tau)$ (or rather on their
belinears). The number of $\psi^A(\tau)$, labeled by the index
$A$, is half the number of $\theta^\alpha(\tau)$ (in $D=2$ $A=1$;
in $D=3$ $A=1,2$ and coincides with the Majorana spinor index
$\alpha$, and in $D=10$ $A=1,\cdots,16$). $\psi^A(\tau)$ are
assumed to transform under a spinor representation of $SO(D-1)$
and should be regarded (like the bosonic worldline variables
$\varphi^i(\tau)$) as non--abelian counterparts of spinorial
coordinates of N D0--branes gauge fixed with $N-1$
$\kappa$--symmetries, the action \p{4} being invariant under a
remaining single $\kappa$--symmetry.

The dependence of the effective mass on $\psi^A$ can be determined
(at least partially) by comparing the ND0--brane action \p{4} with
that of Matrix theory \cite{bfss}, which we perform in Section 4.
Thus up to the second order in $\psi^A$ we have
\begin{equation}\label{4.1}
M(p_i,p_{\varphi},\varphi,\psi)=M_{bos}(p_i,p_{\varphi},\varphi)
+iTr(\psi^A\gamma^i_{AB}[\phi^i,\psi^B])+O(\psi^4)+\cdots \,,
\end{equation}
where $\gamma^i_{AB}$ are $(D-1)$--dimensional gamma matrices and
$M_{bos}(p_i,p_{\varphi},\varphi)$ is the bosonic ND0--brane
effective mass \p{3}.

Before considering the form of the $\kappa$--symmetry
transformations, let us present  global target--space
supersymmetry variations of the fields under which the action
\p{4} is invariant
\begin{equation}\label{5}
\delta_\epsilon\theta^\alpha=\epsilon^\alpha, \qquad\,
\delta_\epsilon
x^M=-i\bar\epsilon\gamma^M\theta-i\bar\epsilon\Gamma\theta{\partial
{M(p_i,p_{\varphi},\varphi,\psi)}\over{\partial{p_i}}}\delta^M_i,
\quad \delta_\epsilon\,p_M=0\,,\qquad  \delta_\epsilon\, e=0\,,
\end{equation}
\begin{equation}\label{6}
\delta_\epsilon\varphi^i=-i\bar\epsilon\Gamma\theta\,
\left[{{\partial M(p_i,p_{\varphi},\varphi,\psi)}\over{N\partial
p_{\varphi^i}}}\right]_{trless}\,, \qquad \delta_\epsilon
p_{\varphi^i}=i\bar\epsilon\Gamma\theta\, \left[{{\partial
M(p_i,p_{\varphi},\varphi,\psi)}\over{\partial
\varphi^i}}\right]_{trless}\,,
\end{equation}
\begin{equation}\label{7}
\delta_\epsilon \psi=i\bar\epsilon\Gamma\theta\, \left[{{\partial
M(p_i,p_{\varphi},\varphi,\psi)}\over{\partial
\psi}}\right]_{trless}\,,
\end{equation}
where $\delta^M_i=1$ when $M=i$, and $\delta^M_i=0$  when $M=0$ or
$M\not =i$.

 Note that since the effective mass  $M(p_{\varphi^i},p_i,
\varphi, \psi)$, is a function of $p_i$, the global supersymmetry
transformation of the spatial coordinates $x^i$ gets modified,
which is the price for the model to be non--invariant under
Lorentz transformations in $D$-dimensional space-time (see
\cite{Sorokin:2001av,partial} for a detailed discussion of this
point).  For similar reasons also the non--abelian $SU(N)$
worldvolume fields $\varphi(\tau)$, $\psi(\tau)$ and the
non--abelian momenta $p_\varphi(\tau)$ non--trivially transform
under target--space supersymmetry.

At this point we should note that $M(p_i, p_{\varphi},\varphi,
\psi)$ itself is {\sl invariant} under the supersymmetry
variations \p{5}--\p{7} and under $\kappa$--variations (see
\p{33}--\p{psi} below) which is crucial for the action \p{4} to be
supersymmetric and $\kappa$--invariant. This is the case for a
generic form of $M(p_{\varphi},p_i, \varphi, \psi)$, the only
requirement being that $M$ does not depend on $x^M$ and
$\theta^\alpha$.

In $D=2$ the effective mass does not depend on $\varphi$ (see
\p{2}). If in the supersymmetric case $M$ does not depend on
$\psi$ the supersymmetric ND0--brane action in $D=2$ possesses
redundant local worldvolume supersymmetries under the variations
$\delta\varphi=2\alpha(\tau)\psi,
\delta\psi=\alpha(\tau)p_\varphi$ with the anticommuting parameter
$\alpha(\tau)$ \cite{Sorokin:2001av}. In higher space--time
dimensions $M$ acquires the dependence on $\varphi$ \p{3}, and the
redundant local worldvolume supersymmetries disappear, even if $M$
does not depend on $\psi$, as  one can directly verify.

The supersymmetry algebra of the transformations \p{5}--\p{7}
generated by the Poisson brackets of the Noether supercharges
derived from the action \p{4} has the following form

\begin{equation}\label{susya}
\{Q_\alpha,Q_\beta\}=2ip_M\gamma^M_{\alpha\beta}
+2iM(p_i,p_{\varphi},\varphi,\psi)\,\Gamma_{\alpha\beta}\,,
\end{equation}
where
\begin{equation}\label{Q}
Q_\alpha=\pi_\alpha+ip_M(\gamma^M\theta)_\alpha +
iM(p_i,p_{\varphi^i},\varphi,\psi)\,(\Gamma\theta)_\alpha\,
\end{equation}
are supercharges and $\pi_\alpha$ are the momenta conjugate to
$\theta^\alpha$. (For the cases of $D=2,3$ and $10$ the explicit
form of $\gamma^M$ and $\Gamma$ has been presented below eq.
\p{4}).

We observe that the superalgebra \p{susya} has the ``central
charge" term proportional to the effective mass $M$ which arises
because the spatial coordinates $x^i$, the $SU(N)$ adjoint
scalars, their momenta and the $SU(N)$ fermions nontrivially
transform under supersymmetry.

 We now present the $\kappa$--symmetry variations of the fields
 of the model under which the action \p{4} is invariant
\begin{equation}\label{33}
\delta_\kappa\theta=\left(p_M\,\gamma^M+M(p_i,p_{\varphi},\varphi,\psi)\,
\Gamma\right)\,\kappa(\tau), \quad \delta_\kappa
x^M=i\delta_\kappa\bar\theta\gamma^M\theta\,
+\,i\delta_\kappa\bar\theta\Gamma\theta\,{{\partial
M}\over{\partial p_i}}\,\delta^M_i\,,
\end{equation}
$$
\delta_\kappa\,p_M=0\,,\qquad \delta_\kappa
e=4i\kappa^\alpha\dot\theta_\alpha\,,
$$
\begin{equation}\label{kf}
\delta_\kappa\varphi^i=i\delta_\kappa\bar\theta\Gamma\theta\,
\left[{{\partial M(p_i,p_{\varphi},\varphi,\psi)}\over{N\partial
p_{\varphi^i}}}\right]_{trless},\quad \delta_\kappa p_{\varphi^i}=
-i\delta_\kappa\bar\theta\Gamma\theta\, \left[{{\partial
M(p_i,p_{\varphi},\varphi,\psi)}\over{\partial
\varphi^i}}\right]_{trless}
\end{equation}
\begin{equation}\label{psi}
\delta_\kappa\psi=-i\delta_\kappa\bar\theta\Gamma\theta\,
\left[{{\partial M(p_i,p_{\varphi},\varphi,\psi)}\over{\partial
\psi}}\right]_{trless}\,,
\end{equation}
where $\kappa^\alpha(\tau)$ is the local fermionic parameter. Let
us stress again that the effective mass is invariant under the
$\kappa$--symmetry variations as well as under the target space
supersymmetry.

\section{Comparison with Matrix Theory}
In an certain limit of type IIA string theory the non--abelian DBI
action for ND0--branes is known to reduce to the Matrix theory
action \cite{bfss}
\begin{equation}\label{5.1}
S={m}\,Tr \, \int\, d\tau\,\left({1\over 2}\,
\dot\Phi^i\dot\Phi^j-{1\over
4}[\Phi^i,\,\Phi^j]\,[\Phi^i,\,\Phi^j]+i\Theta^A\dot\Theta^A-
i\Theta^A\gamma^i_{AB}[\Phi^i,\,\Theta^B] \right),
\end{equation}
where $\Phi^i(\tau)=x^i\,I+\phi^i$ are the ND0--brane
$U(N)$-valued scalar fields and $\Theta^A$ are $U(N)$-valued
fields transforming under a spinor representation of $SO(D-1)$.

To reduce the ND0--brane action \p{4} to the Matrix theory action
\p{5.1} and to relate $\theta^\alpha$ and $\psi^A$ to $\Theta^A$
we should fix the worldvolume diffeomorphisms and the $\kappa$
symmetry by imposing a static gauge
\begin{equation}\label{5.2}
x^0=\tau,\quad \theta^2_\alpha=0~~{\rm in}~D=3\quad {\rm and}\quad
\theta^\alpha=(\gamma^{11}\theta)^\alpha ~~~{\rm in}~D=2,10
\end{equation}
so that, for example in $D=10$, 32--component $\theta^\alpha$
reduces to a 16--component Majorana--Weyl spinor $\theta^A$ which
together with $\psi^A$ form the $U(N)$ spinor $\Theta^A=\theta^A
\,I + \psi^A$.

Then we keep in the action \p{4} the terms up to the second order
in the momenta $p_{\Phi^i}$ and in $[\Phi^i,\,\Phi^j]$, integrate
over $p_0$ and $p_{\Phi^i}$ and skip all derivative terms except
for the kinetic terms. We thus arrive at the matrix theory action
\p{5.1}. Note that the form of the last term in \p{5.1} has
prompted us the dependence of the effective mass \p{4.1} on the
non--abelian fermions $\psi^A$.

\section{The Lorentz--covariant super--ND0--brane system} As it
has been considered in detail in \cite{Sorokin:2001av,partial} the
ND--brane actions proposed in
\cite{Tseytlin:1999dj,Myers:1999ps,Taylor:2000pr} are not Lorentz
(or diffeomorphism) invariant in D--dimensional target space
except for the space filling branes. This is not because they are
constructed in the static gauge, which can be removed by restoring
the reparametrization invariance in the way discussed above. The
main reason is that the center--of--mass $U(1)$ coordinates $x^i$
of the ND--brane get mixed with the non--abelian $SU(N)$ scalar
fields $\varphi^i$, which follows from the form of the action
\p{3.1}. This implies that the motion of the center of mass of the
ND--brane as a whole depends on the internal excitations inside
the system, which is rather strange. One may wonder whether the
computation of string amplitudes associated with N coincident
D-branes of higher codimension confirms such an effect.

 We now consider the Lorentz--covariant counterpart of the above model.
 For this we assume the effective mass in (\ref{4}) to be independent of the
spatial momenta $p_i$. This ensures the free motion of the
ND0--brane center of mass. If so, we can now assume that the
indices $i$ and $A$ of $\varphi^i(\tau)$, $p_{\varphi^i}$ and
$\psi^A$ are the indices of a corresponding representation of an
{\it independet} internal group $SO(D-1)$ which {\it a priori} is
not related to the spatial subgroup of the Lorentz group
$SO(1,D-1)$.

The supersymmetric and $\kappa$ invariant action for the Lorentz
invariant ND0--brane is
\begin{eqnarray}\label{36}
S&=&\int d\tau {\rm Tr}\left\{{1\over N}p_M(\dot
x^M+i\theta\gamma^M\dot\theta)\right. \left.-{{e(\tau)}\over
2N}\left[p_Mp^M+M^2(p_\varphi,\varphi,\psi)\right]
\right.\nonumber \\
&& \left.+{i\over
N}\,M(p_\varphi,\varphi,\psi)\, \bar\theta\Gamma\dot\theta
+p_\varphi\dot\varphi+i\bar\psi\dot\psi \right\}\,.
\end{eqnarray}
The variation properties of the fields with respect to the
symmetries of the action \p{36} are the same as written in
equations \p{5}--\p{7}, \p{33}--\p{psi}, except that now the
spatial coordinates $x^i$ transform in the standard way under the
$\kappa$--symmetry and target--space supersymmetry, i.e. there is
no contribution of the effective mass into their variation, and
the variation of $x^M=(x^0,x^i)$ is Lorentz invariant.

\section{Conclusion and discussion}
We have constructed the action for N coincident D0--branes which
is target--space supersymmtric and invariant under local worldline
fermionic $\kappa$--symmetry in a D-dimensional $N=2$ superspace,
particular cases being $D=2,3$ and type IIA $D=10$. The action has
a generic structure determined by a super-- and
$\kappa$--invariant effective mass which is a generic function of
the non--abelian $SU(N)$ fields of the model and their momenta,
and which also depends on the spatial momentum of the center of
mass of the system. It is crucial for the invariance of the action
that the effective mass does not depend on $x^M$ and
$\theta^\alpha$. In the bosonic limit the explicit form of the
effective mass is dictated by the non--abelian DBI structure of
the ND0--brane action, and it should still to be determined for
$D=10$.

We have compared the supersymmetric ND0--brane action with that of
Matrix theory, which has allowed us to fix the dependence of the
effective mass \p{4.1} on the non--abelian $SU(N)$ fermions
$\psi^A$ up to the second order. A problem which remains is to
determine higher order fermionic terms in the effective mass.

It would be of interest to apply the T--duality procedure to the
supersymmetric ND0--brane model for getting supersymmetric actions
for higher dimensional coincident D--branes.

It would be also interesting to compare the above ND0--brane
construction with that of \cite{dhl} for the space filling
supersymmetric ND2--brane by performing the worldvolume
dimensional reduction of the latter.

\section*{Acknowledgements}
D.S. is grateful to I. Bandos, J. de Azcarraga, J. Gomis and M.
Plyushchay for interest to this work and useful discussions. Work
of D.S. was partially supported by the European Commission TMR
Programme HPRN-CT-2000-00131 to which the author is associated
with the University of Padua, by the Grant N 383 of the Ukrainian
State Fund for Fundamental Research and by the INTAS Research
Project N 2000-254.


\begin{thebibliography}{99}

\bibitem{Tseytlin:1999dj}
A.~A.~Tseytlin, ``Born-Infeld action, supersymmetry and string
theory,'' hep-th/9908105.
%%CITATION = HEP-TH 9908105;%%
\bibitem{Myers:1999ps}
R.~C.~Myers,
%``Dielectric-branes,''
JHEP {\bf 9912} (1999) 022 [hep-th/9910053];
%%CITATION = HEP-TH 9910053;%%\\
``Nonabelian D--Branes and Noncommutative geometry",
hep--th/0106178.
%%CITATION = HEP-TH 0106178;%%
\bibitem{Taylor:2000pr}
W.~I.~Taylor and M.~Van Raamsdonk, Nucl.\ Phys.\ B {\bf 558}
(1999) 63 [hep-th/9904095];
%%CITATION = HEP-TH 9904095;%%
%``Multiple Dp-branes in weak background fields,''
Nucl.\ Phys.\ B {\bf 573} (2000) 703 [hep-th/9910052].
%%CITATION = HEP-TH 9910052;%%
\bibitem{roo}
A.~Collinucci, M.~de Roo and M.~G.~Eenink, ``Derivative
corrections in 10-dimensional super-Maxwell theory,''
hep-th/0212012.
%%CITATION = HEP-TH 0212012;%%
\bibitem{Sorokin:2001av}
D.~Sorokin,
%``Coincident (super)-Dp-branes of codimension one,''
JHEP {\bf 0108} (2001) 022 [hep-th/0106212].
%%CITATION = HEP-TH 0106212;%%
\bibitem{partial}
D.~Sorokin,
%``Space-Time Symmetries And Supersymmetry Of Coincident D-Branes,''
Fortsch.\ Phys.\  {\bf 50} (2002) 973.
%%CITATION = FPYKA,50,973;%%
\\
D.~Sorokin, ``On superbranes with non-Abelian worldvolume
fields,'' in ``Supersymmetries and Quantum Symmetries" E. Ivanov
et. al. Editors, Dubna 2002, pp. 74--85.
\bibitem{dhl}
J.~M.~Drummond, P.~S.~Howe and U.~Lindstrom,
%``Kappa-symmetric
%non-Abelian Born-Infeld actions in three dimensions,''
Class.\ Quant.\ Grav.\  {\bf 19} (2002) 6477  [hep-th/0206148].
%%CITATION = HEP-TH 0206148;%%
\bibitem{super}
D.~P.~Sorokin,
%``Superbranes and superembeddings,''
Phys.\ Rept.\  {\bf 329} (2000) 1 [hep-th/9906142].
%%CITATION = HEP-TH 9906142;%%
\bibitem{b}
E.~A.~Bergshoeff, M.~de Roo and A.~Sevrin,
%``Towards a supersymmetric non-Abelian Born-Infeld theory,''
Int.\ J.\ Mod.\ Phys.\ A {\bf 16} (2001) 750 [hep-th/0010151];
%%CITATION = HEP-TH 0010151;%%
``Non-Abelian Born-Infeld and kappa-symmetry,'' hep-th/0011018;
%%CITATION = HEP-TH 0011018;%%
Fortsch.\ Phys.\  {\bf 49} (2001) 433 [hep-th/0011264].
%%CITATION = HEP-TH 0011264;%%
\bibitem{pp}
S.~S.~Pal and S.~Panda, ``Coincident Dp-branes in codimension
two,'' hep-th/0211115.
%%CITATION = HEP-TH 0211115;%%
\bibitem{bfss}
T.~Banks, W.~Fischler, S.~H.~Shenker and L.~Susskind,
%``M theory as a matrix model: A conjecture,''
Phys.\ Rev.\ D {\bf 55} (1997) 5112 [hep-th/9610043].
%%CITATION = HEP-TH 9610043;%%


\end{thebibliography}
\end{document}